\documentclass[twocolumn,trackchanges]{aastex701}
\usepackage{bm}
\usepackage{amsmath}
\usepackage{enumitem}
\usepackage{graphicx}
\usepackage{caption}
\usepackage{hyperref}
\usepackage{mathcomp}

\begin{document}

% Hotta: Genetic Algorithm for Inferring Model Parameters for Flux Transport Dynamo Simulationが良いかなと思います。(済)
\title{Genetic Algorithm for Inferring Model Parameters for Flux Transport Dynamo Simulation}

\author[orcid=0000-0000-0000-0001]{Y. Shimizu}
\affiliation{Institute for Space-Earth Environmental Research, Nagoya University, Furo-cho, Chikusa-ku, Nagoya, Aichi 464-8601, Japan}
\email[show]{shimizu.yuya.h3@s.mail.nagoya-u.ac.jp}  

%Hotta 堀田のはEメールアドレスを示さない方がありがたいです。（済）
\author[orcid=0000-0002-6312-7944]{H. Hotta}
\affiliation{Institute for Space-Earth Environmental Research, Nagoya University, Furo-cho, Chikusa-ku, Nagoya, Aichi 464-8601, Japan}
\email[show]{}

\begin{abstract}
%レフェリー指示により削除 In this work, we introduce a method for automatically inferring the mean-field solar dynamo parameters using a genetic algorithm. 
The Sun exhibits an 11-year cyclic variation, maintained by dynamo action in the solar interior. Mean-field flux transport dynamo models have successfully reproduced most of the features observed in solar cycles, while the model includes many free parameters, such as the speed of the meridional flow and the amplitude of the poloidal field generation. Inferring these free parameters is on demand because they correspond to the solar interior condition. We suggest a novel method for inferring the free parameters using a genetic algorithm. At each generation, we evaluate the fitness of our simulation against the observational data and optimize the parameters. 
We apply our method to the observed solar cycle data from 1723 to 2024 and successfully reproduce the observations from both qualitative and quantitative perspectives. We expect our method to be applicable to sunspot numbers, even those obtained from isotope data and historical documents, in the future, to better understand past solar interior dynamics.
\end{abstract}


\section{introduction} \label{section:introduction}
%TODO この分野の重要性を明示
The number of sunspots varies with the 11-year cycle. The cyclic behavior exhibits many rules, such as equatorward migration of sunspot latitude \citep{maunder_1904MNRAS..64..747M}, and a symmetric rule for polarity of sunspot pairs  about the equator \citep{hale_1919ApJ....49..153H}. Elucidating these features is one of the most important research topics in solar physics \citep[see review by][]{charbonneau_2020LRSP...17....4C}.
% イントロのこんな最初の方に、この研究の目的を書かない This study focuses on the $11$ year cycle of sunspot number fluctuations, and there are irregularities in the observational data. 
In addition to the regular cyclic behavior, it is known that the cycle amplitude and period depend on the individual cycle \citep{hathaway_2010LRSP....7....1H}. 
% The height of the sunspot number peaks and the length of the period vary with each cycle. 
The cycle and the longer-scale variation can be attributed to the magnetic field generation, i.e., the dynamo action, in the solar interior \citep{Karak2023}.
% and are also considered to be a factor in the grand minima, when sunspots almost disappear and activity levels are strongly reduced \citep{1976Sci...192.1189E,1993A&A...276..549R}. 
\par
%TODO 背景知識としてさまざまな研究のアプローチを紹介、研究意義に繋げる
Numerical simulations play a key role in addressing the origin and the variation of the sunspot cycle, as direct observation of the solar deeper layers, where the dynamo action is thought to occur, is extremely limited. Numerical simulation enables us to examine the solar interior dynamics and are considered an effective solution for investigating solar activity. The flux transport dynamo is one of the most promising models for reproducing the solar cycle and its features \citep{choudhuri_1995A&A...303L..29C, 1999ApJ...518..508D}. In the model, the pre-existing poloidal field is stretched by differential rotation to produce the toroidal field around the base of the convection zone \citep[$\Omega$-effect:][]{1955ApJ...122..293P}. Then the toroidal field rises to the surface due to magnetic buoyancy and/or thermal convection \citep{fan_2021LRSP...18....5F}. During the flux rise, the Coriolis force bends the field to produce a poloidal magnetic field around the surface \citep[so called the Babcock-Leighton $\alpha$-effect:][]{1961ApJ...133..572B,1969ApJ...156....1L}. The generated poloidal field is transported by the meridional flow and turbulent diffusion \citep{hotta_2010ApJ...709.1009H}. 
The flux transport dynamo simulation is a mean-field model that includes several free parameters, such as the meridional flow speed and the amplitude of the Babcock--Leighton $\alpha$-effect. There have been several attempts to infer the free parameters and reproduce the observational feature of the sunspot cycle. For example, \cite{2017ApJ...834..133L} constructed a solar dynamo model optimized for cycle 21, in which a genetic algorithm tuned the input free parameters to match the observations, and the model was then applied to reproduce solar cycles and explore long-term variability. \cite{2010ApJ...724.1021K} evaluate time variations in the meridional flow speed manually to reproduce changes in cycle period and amplitude, as well as the Maunder Minimum.
% The magnetic flux transport dynamo used by them and us provides a good explanation for the solar cycle scenario.
% First, consider the existence of a magnetic field connecting the poles. This magnetic field is affected by differential rotation at the tachocline, generating a strong toroidal magnetic field \citep{1955ApJ...122..293P}. 
% Magnetic buoyancy acts on the magnetic field, transporting it to the surface to form sunspots. At this point, the magnetic flux is twisted due to the Coriolis force, generating pairs of tilted sunspots \citep{1993A&A...272..621D}.
% As sunspots decay, the poloidal magnetic field components increases on the solar surface \citep{1961ApJ...133..572B,1969ApJ...156....1L}. These are transported toward the poles by meridional flow, gradually weakening and reversing the polar magnetic field. And it generates a new magnetic field connecting the poles.
% These factors affecting the magnetic field —meridional flow, differential rotation, magnetic buoyancy, the $\alpha$ effect, and others— are crucial for explaining irregularities, making simulation studies to understand the interior essential.\\
\par
In the observational aspect, isotope analyses and historical documents research to understand the past solar activity have significantly progressed in the past two decades. They are studying solar activity from an ancient era for which no precise records are available.
For example, \cite{2004SoPh..224..317M} carried out a high-precision analysis of carbon-$14$ in annual ring data during the Maunder Minimum and suggested that even during the grand minimum, the dynamo may not have ceased and the solar polarity reversal may have been maintained.
\cite{2025arXiv250214665H} presents an example of research that utilizes historical documents from the 17th to 18th centuries to reconstruct long-term solar variability, providing fundamental data sets for the reconstruction of ancient solar activity and further supporting the hemispheric asymmetry during the Maunder Minimum.
\par
Considering the recent rapid accumulation of the sunspot and solar activity data, an automatic method(s) to evaluate the solar internal dynamics is in high demand. \cite{2017ApJ...834..133L} adapted a similar approach, but they obtained constant parameters in time. In reality, we expect most things to depend on time, and we try to infer the temporal variation of parameters in this study.
\par
This paper is structured as follows: The methods for the numerical simulation and the genetic algorithm are explained in Section \ref{sec:method}. The results for validating our method and the application to the observational data are shown in \ref{sec:result}.

\section{method} \label{sec:method}
\subsection{Numerical model}
We solve the axisymmetric induction equation in spherical coordinates $(r,\theta, \phi)$ with the kinematic assumption, i.e., the flow field $\bm{u}_\mathrm{p}$ is given. 
The magnetic field $\bm{B}$ is decomposed to the poloidal mean field $\bm{B}_\mathrm{p} = \nabla \times [A_\phi(r,\theta)\boldsymbol{e_\phi}]$ 
and the toroidal field $B_\phi\bm{e}_\phi$, and our equations can be written as \citep[e.g., ][]{1999ApJ...518..508D}

\begin{subequations} \label{eq:induction}
\begin{align}
% Poloidal field
\frac{\partial A_\phi}{\partial t}
&+ \frac{1}{r\sin\theta}(\bm{u}_\mathrm{p} \cdot \nabla)(r\sin\theta A_\phi) \notag \\
&= \eta \left(\nabla^2 - \frac{1}{r^2\sin^2\theta} \right) A_\phi 
+ S(r,\theta;B_\phi), \label{eq:induction_a} \\[1ex]
%
% Toroidal field
\frac{\partial B_\phi}{\partial t}
&+ \frac{1}{r} \left[
\frac{\partial}{\partial r}(r u_r B_\phi)
+ \frac{\partial}{\partial \theta}(u_\theta B_\phi)
\right] \notag \\
&= r \sin\theta (\boldsymbol{B_\mathrm{p}} \cdot \nabla) \Omega 
- \frac{1}{r} \frac{\partial \eta}{\partial r} \frac{\partial}{\partial r}(r B_\phi) \notag \\
&\quad + \eta \left( \nabla^2 - \frac{1}{r^2 \sin^2\theta} \right) B_\phi, \label{eq:induction_b}
\end{align}
\end{subequations}
where $\bm{u}_\mathrm{p}$ represents the meridional flow velocity ($\bm{u}_\mathrm{p}=u_r\bm{e}_r+ u_\theta\bm{e}_\theta$), 
$\Omega$ is the angular velocity representing the differential 
rotation, and $S(r,\theta;B_\phi)$ shows the Babcock-Leighton $\alpha$-effect source term. When the magnetic field is generated around the base of the convection zone, the corresponding magnetic field is generated around the surface.
% Without this term, the magnetic field cannot be sustained under the axisymmetric assumption \citep{1953sun..book..532C}. 
$\eta$ represents turbulent diffusivity. In this study, we basically follow the settings of \cite{2008A&A...483..949J} (see their equations (7) and (8)). While the compression term $-\bm{B}\nabla\cdot\bm{u}$ is ignored in their study, we do not do so.
\par

%%%%%%%%%%%%%%%%%%%%%%%%%%%%%%%%%%%%%%%%%%%%%%%%%%%%%%%%%%%%%%%%%%%%%%%%%%%%%%%%%%%%%%%%%%%%%%%%%%%%%%%%%%%
%計算領域
Equations \eqref{eq:induction} are solved on the meridional plane $(0\leq\theta\leq\pi)$. The radial domain extends $0.65R_\odot < r < R_\odot$, where $R_\odot=6.96\times 10^{10}~\mathrm{cm}$ is the solar radius.
\par
%初期条件
% Hotta: Initial conditionについてはここでは何も書かなくて良いかな
% Initial conditions do not remain constant throughout but vary within the methodology. Details are described in section \ref{section:Initial condition}.
%境界条件
The bottom boundary at $r=0.65R_\odot$ is set to be a perfect conductor as
\begin{equation} \label{eq:bc_bottom}
A_\phi = 0 \quad\text{and}\quad \frac{\partial (r B_\phi)}{\partial r} = 0.
\end{equation}
The upper boundary condition at $r=R_\odot$: we smoothly match the magnetic field to an external potential field, i.e., we assume that the exterior of the Sun is a vacuum and current-free. This condition is expressed as
\begin{equation} \label{eq:bc_upper}
\left( \nabla^2 - \frac{1}{r^2\sin^2\theta}\right)A_\phi = 0 ~\text{and}~B_\phi = 0.
\end{equation}
At $\theta=0$ and $\theta=\pi$ boundaries, we impose the regularity conditions as
\begin{equation} \label{eq:bc_polar}
A_\phi = 0~\text{and}~B_\phi = 0.
\end{equation}
We employ the second-order space-centered derivative and second-order Total Variation Diminishing (TVD) Runge-Kutta method for time integration. For details of TVD Runge-Kutta method, please refer to the appendix. No artificial viscosity is included. The calculation is stabilized with the turbulent diffusivity.
Our simulation is consistent with the results of a solar mean field dynamo benchmark test \citep{2008A&A...483..949J}.
We use the formulations introduced in \cite{2008A&A...483..949J} for the following ingredients of the meridional flow, the $\alpha$-effect source term, and the turbulent diffusivity.
\par
%%%%%%%%%%%%%%%%%%%%%%%%%%%%%%%%%%%%%%%%%%%%%%%%%%%%%%%%%%%%%%%%%%%%%%%%%%%%%%%%%%%%%%%%%%%%%%%%%%%%%%%%%%%
The profile of differential rotation $\Omega$ is defined as
\begin{equation} \label{eq:omega}
\Omega (r,\theta) = \Omega_\mathrm{c} + \frac{1}{2}\left[1 + \mathrm{erf} \left(\frac{r - r_\mathrm{c}}{d}\right)\right]\left(\Omega_\mathrm{e} - \Omega_\mathrm{c} - c_2\cos^2\theta\right),
\end{equation}
where $\Omega_\mathrm{e} /2\pi=460\mathrm{nHz}$, $\Omega_\mathrm{c} = 0.92 \Omega_e$, $c_2=0.2\Omega_\mathrm{e}$, $r_\mathrm{c}=0.7R_\odot$, $d=0.02R_\odot$. 
In our expression, the core rotates with $\Omega_\mathrm{c}$ and the thickness of the tachocline at $r=r_\mathrm{c}$ is $d$. This angular velocity distribution roughly reproduces the observational result by helioseismology \citep{schou_1998ApJ...505..390S,hotta_2023SSRv..219...77H}.
\par
% Next we describe the meridional flow. Meridional flow is a flow field that exists 
% on and within the Sun.
% The meridional flow plays a role in advection of 
% the magnetic field in the solar interior. 
The profile of meridional flow is given as 
%mod1 上記に式の形はJouveを参照すると書いているのでいらないと判断\citep{2008A&A...483..949J}
\begin{subequations} \label{eq:meridional_flow}
\begin{align}
u_r(r,\theta)
  &= -u_0\frac{2(R_\odot - r_\mathrm{b})}{\pi r}\frac{(r-r_\mathrm{b})^2}{(R_\odot-r_\mathrm{b})^2}\notag\\
  &\times\sin\left(\pi\frac{r-r_\mathrm{b}}{R_\odot-r_\mathrm{b}}\right)(3\cos^2\theta-1), \\[1ex]
u_\theta(r,\theta)
  &= u_0\left[\frac{3r-r_\mathrm{b}}{R_\odot - r_\mathrm{b}}\sin\left(\pi\frac{r-r_\mathrm{b}}{R_\odot-r_\mathrm{b}}\right)\right.\notag\\
  &\quad\left.+\frac{r\pi}{R_\odot-r_\mathrm{b}}\frac{r-r_\mathrm{b}}{R_\odot-r_\mathrm{b}}
     \cos\left(\pi\frac{r-r_\mathrm{b}}{R_\odot-r_\mathrm{b}}\right)\right]\notag\\
  &\times\frac{2(R_\odot-r_\mathrm{b})}{\pi r}\frac{r-r_\mathrm{b}}{R_\odot-r_\mathrm{b}}\cos\theta\sin\theta ,
\label{eq:meridional_b}
\end{align}
\end{subequations}
where $r_\mathrm{b}=0.65R_\odot$ and the amplitude of the meridional flow $u_0$ is a parameter to be varied and inferred in this study. 
The flow is set to vanish at $r=0.65R_\odot$.
% This is consistent with the Waldmeier effect\citep{1935MiZur..14..105W}.
\par
With the Babcock--Leighton $\alpha$ effect, the source term is defined as
\begin{align} \label{eq:alpha}
& S (r,\theta;B_\phi) = \frac{s_0}{2}\left[1 + \mathrm{erf} \left(\frac{r - r_1}{d_1}\right)\right]\left[1-\mathrm{erf}\left(\frac{r-R_\odot}{d_1}\right)\right]\notag\\
&\times\left[1+\left(\frac{B_\phi(r_\mathrm{c},\theta,t)}{B_0}\right)^2\right]^{-1}\cos\theta\sin\theta B_\phi(r_\mathrm{c},\theta,t),
\end{align}
where $r_1=0.95R_\odot$, $d_1=0.01R_\odot$. The amplitude of the source term $s_0$ is another parameter to be varied and inferred in this study. To prevent the magnetic field from growing exponentially without bounds, we introduce a nonlinear quenching term $[1+\left(B_\phi(r_c,\theta,t)/B_0\right)^2]^{-1}$ with $B_0$ being an arbitrary normalization constant.
In this paper, $B_\phi$ and $A_\phi$ are shown as values normalized by $B_0$.
We note some previous research for the tilt and the latitude of sunspots support the existence of the quenching in the Babcock-Leighton process \citep[e.g.][]{Jha_2020, 2020ApJ...900...19J, Dey_2025}.
\par
Here, we emulate the observed cycle-to-cycle variation by adjusting $u_0$ and $s_0$, because the prevailing theory shows that the irregularity of the solar cycle comes from the fluctuations in the Babcock-Leighton $\alpha$ effect and the meridional circulation \citep{1992A&A...253..277C,2009RAA.....9..953C,2013IAUS..294..433K}. 
Thus, we consider the time variations of the meridional flow speed $u_0$ and the source term amplitude $s_0$. 
A sine-series expansion controls the time dependence of these parameters as
%mod1 上付き文字はイタリックに直す
%mod Teなどの説明を加える
\begin{subequations} \label{eq:time_variation}
\begin{align}
u_0(t)
&= \frac{a^u_\mathrm{s}(T_\mathrm{e}-t)+a^u_\mathrm{e}(t-T_\mathrm{s})}{T_\mathrm{e}-T_\mathrm{s}} + \sum_{l=1}^{l_\mathrm{max}} a^u_l \sin l\omega t, \\
s_0(t)
&= \frac{a^s_\mathrm{s}(T_\mathrm{e}-t)+a^s_\mathrm{e}(t-T_\mathrm{s})}{T_\mathrm{e}-T_\mathrm{s}} + \sum_{l=1}^{l_\mathrm{max}} a^s_l \sin l\omega t.
\label{eq:induction_b}
\end{align}
\end{subequations}

where $T_\mathrm{s}$ and $T_\mathrm{e}$ are the start and end time of inference. We set $l_\mathrm{max}=4$ to compromise expressiveness and inversion difficulty.
Note that $l=3$ is excluded to determine the start and end points of the parameter's time variation by $a_\mathrm{s}^u,~a_\mathrm{e}^u,~a_\mathrm{s}^s~\mathrm{and}~a_\mathrm{e}^s$.

The angular frequency $\omega$ is defined as 
\begin{align}
    \omega = \frac{2\pi}{2(T_\mathrm{e}-T_\mathrm{s})},
\end{align}

% By increasing the order n, complex time variations can be expressed. Here we set $n=3$, 
% because we find it to be empirically sufficient. 
% Hotta これもあとで言えば良いことと思います。
% By appropriately adjusting the coefficient $a_n$, we can reproduce a solar cycle close to the observation and obtain 
% the time variation of the dynamo parameters $u_0$ and $s_0$.\\
\par
The profile of turbulent diffusivity is given as
\begin{equation} \label{eq:diffusivity}
\eta (r) = \eta_\mathrm{c} + \frac{1}{2}(\eta_\mathrm{t} - \eta_\mathrm{c})\left[1 + \text{erf}\left(\frac{r-r_\mathrm{c}}{d}\right)\right],
\end{equation}

where $\eta_\mathrm{c} = 1\times 10^{9}\mathrm{~cm^2~s^{-1}}$, $\eta_t = 1\times 10^{11}\mathrm{~cm^2~s^{-1}}$.
We assume that the diffusivity in the convection zone is dominated by $\eta_t$ whereas below the tachocline, turbulence rapidly weakens and diffusion becomes about $\eta_c$.
%%%%%%%%%%%%%%%%%%%%%%%%%%%%%%%%%%%%%%%%%%%%%%%%%%%%%%%%%%%%%%%%%%%%%%%%%%%%%%%%%%%%%%%%%%%%%%%%%%%%%%%%
\subsection{Genetic algorithm(GA)} \label{section:GA procedures}
We use Genetic Algorithm (GA) to obtain the time variation of the parameters $s_0(t)$ and $u_0(t)$, i.e., coefficients in equation \eqref{eq:time_variation}.
%%%%%%%%%%%%%%%%%%%%%%%%%%%%%%%%%%%%%%%%%%%%%%%%%%%%%%%%%%%%%%%%%%%%%%%%%%%%%%%%%%%%%%%%%%%%%%%%%%%%%%%%

%%%%%%%%%%%%%%%%%%
\begin{figure*}[t]
  \centering
  \includegraphics[width=\textwidth]{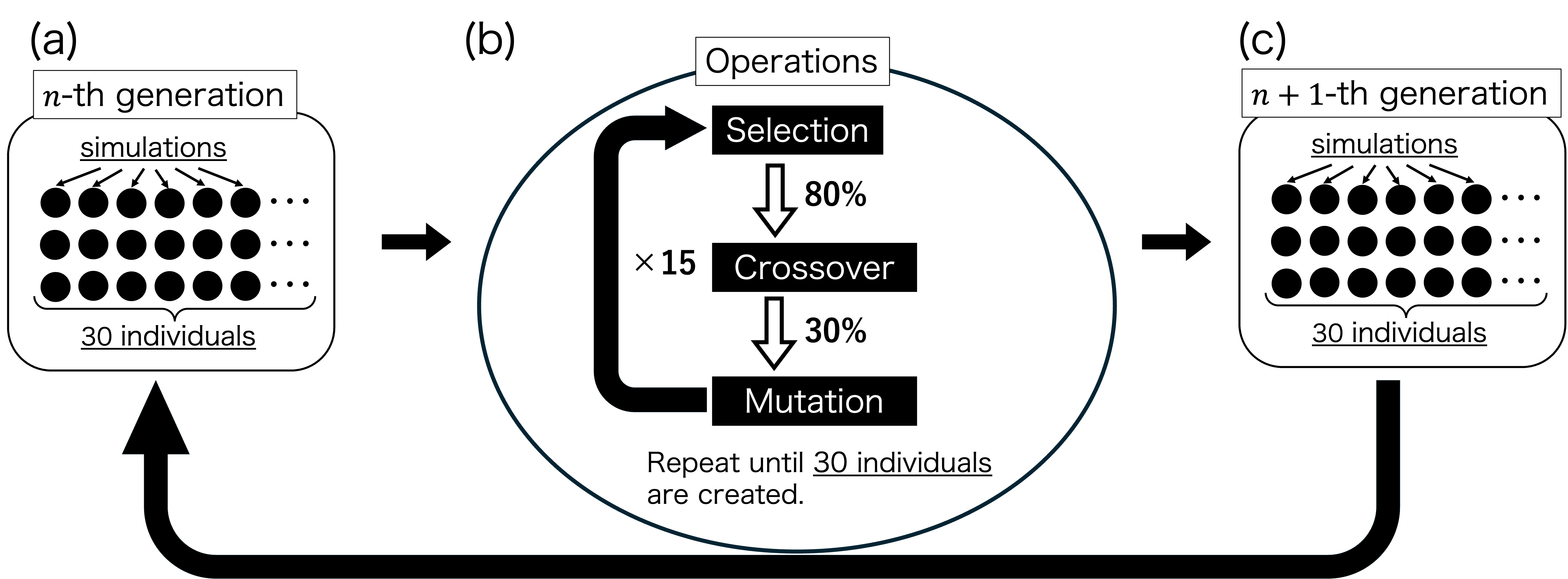}
  \caption{Overview of our Genetic Algorithms. Panel(a) represents $n$-th generation. The black filled circles indicate individuals. Panel (b) represents the Genetic Algorithms operations. After two individuals are selected, crossover and mutation are performed with $80\%$ and $30\%$ probability, respectively. We repeat these operations until 30 individuals are created. Panel(c) represents the $n+1$-th generation. By repeating (a) to (c), leads to the optimal solution.}
  % Red circles indicate individuals with high fitness parameters, which are selected preferentially. Panel b represents the two selected individuals. Panel c represents individuals who undergo crossover with an $80\%$ probability. Panel d represents individuals that undergo a mutation with a $30\%$ probability.
  % By repeating operations (b), (c), and (d), the $(n+1)$-th generation represented in panel e is generated.
  % Then, by repeating operations (a) to (e), optimization progresses, leading to the optimal solution represented in panel f.
  \label{figure_1}
\end{figure*}
%%%%%%%%%%%%%%%%%%
We first randomly prepare 30 tuples of coefficients $a_n$ (Figure \ref{figure_1}a). Each tuple is called an ``individual.'' The group of individuals is referred to as the ``population.'' The initial population generated in this way is referred to as the 0-th generation. Our defined fitness parameter evaluates each individual. We use time series sunspot number (SSN) data $S_\mathrm{GT}(t_i)$ as a measure of solar cycles. The fitness parameter is calculated with the observed/simulated ground truth data $S_\mathrm{GT}(t_i)$ and simulated data $S_\mathrm{sim}(t_i)$. 
We define the fitness parameter as
%Simizu e_\mathrm{sunspot}と誤差を置いた
\begin{equation} \label{eq:fitness}
  F = \alpha r_\mathrm{sunspot} - (1-\alpha) e_\mathrm{sunspot},
\end{equation}
where the weight is $\alpha=0.9$. 
$r_{\text{sunspot}}$ is the correlation coefficient given as
\begin{equation} \label{eq:correlation coefficient}
    r_\mathrm{sunspot} = \frac{\sum_{i} S'_\mathrm{GT}(t_i)S'_\mathrm{sim}(t_i)}{\sqrt{\sum_{i} S'^2_\mathrm{GT}(t_i)\sum_{i}S'^2_\mathrm{sim}(t_i)}},
\end{equation}
where $\langle S_*\rangle$ denote the ensemble average and $S_*'=S_* - \langle S_*\rangle$.
$e_\mathrm{sunspot}$ is the normalized mean square error given as 
\begin{equation} \label{eq:fitness_erorr}
    e_\mathrm{sunspot}=\frac{\sum_{i}\left[S_\mathrm{GT}(t_i)-S_\mathrm{sim}(t_i)\right]^2}{\sum_{i}S_\mathrm{GT}^2(t_i)}.
\end{equation}
%Shimuzu 以下のように平均の説明も書くべきか。標準偏差の説明にも同様な説明がすでにある。
% $\langle S_\mathrm{GT}(t_i)\rangle~\mathrm{and}~\langle S_\mathrm{sim}(t_i)\rangle$ are the ensemble average.
% We note that SSN is merely an indicator of solar activity, so we intend to reproduce the overall variation rather than the short-term variation.
%Shimizu ここに黒点定義を書く
As for SSN, we adopt the square of the toroidal field ($B_\phi^2$) at latitude $15\tcdegree$ at the base of the convection zone \citep{1999ApJ...518..508D} defined as:
\begin{equation}\label{eq:def_SSN}
    \mathrm{SSN} = \gamma B^2_\phi(r=0.7R_\odot,~\theta=15\tcdegree),
\end{equation}
where the coefficient is $\gamma=5.865$ in this study.

\par
Then, we generate the next generation population, where GA undergoes three genetic operators: ``selection,'' ``crossover,'' and ``mutation'' (Figure \ref{figure_1}b). These three operations are regarded as a ``regular set.'' A regular set operation generates two individuals for the next generation (Figure \ref{figure_1}c). We repeat the regular set to generate 30 next-generation individuals.
%TODO 上の保険いらんかも？？
%TODO 周期の相関係数を使う場合もあることをあとで追記
%%%%%%%%%%%%%%%%%%%%%%%%%%%%%%%%%%%%%%%%%%%%%%%%%%%%%%%%%%%%%%%%%%%%%%%%%%%%%%%%%%%%%%%%%%%%%%%%%%%%%%%%
We explain the actual operations of selection, crossover, and mutation, and the methods employed in this study.
\par
Selection is an operator that chooses two individuals as parents for the next generation. We select individuals based on our fitness parameter $F$. We adopt an adaptive selection pressure scheme, which dynamically adjusts the strength of selection pressure, i.e., how strongly better-adapted individuals are favored, according to the progress of GA \citep{1688348}. We use tournament selection. In this method, $m$ individuals are randomly chosen from the population, and the top two are selected as parents. The tournament size $m$ is determined by the selection pressure. The selection pressure controls both ``progress'' and ``diversity'' of the optimization. We evaluate the progress of GA using the relative change rate in the best fitness parameter $\Delta F_n$ and diversity $D_n$ as indicators. $\Delta F_n$ is defined as
\begin{equation} \label{eq:change rate}
\Delta F_n = \frac{|F_{n} - F_{n-1}|}{|F_{n-1}|},
\end{equation}
where $F_n$ represents the best fitness parameter $F$ of the $n$-th generation. We use the mean of the standard deviations of all parameters after normalization as a diversity measure.
Each parameter is 
%mod1 調べても第一正規化？min-max-normalizationではない？？
first normalized by its range,
\begin{equation} \label{eq:normalized}
\tilde{x}^{(k)}_{i,n} = \frac{x^{(k)}_{i,n} - \min\left(x^{(k)}_{i,n}\right)}{\max\left(x^{(k)}_{i,n}\right) - \min\left(x^{(k)}_{i,n}\right)},
\end{equation}
where $x^{(k)}_{i,n}$ is a parameter for the inference with the parameter kind $k$, i.e., $a^s_0$, $a^u_0$, $a^s_1$, and $a^u_1$...,  individual identifier $i$ and the generation $n$.
Then our diverse measure is defined as
% Shimizu 以下の式であっています。codeではnp.std()を使用しており、(https://numpy.org/doc/2.1/reference/generated/numpy.std.html)で正しいことを確認済み
\begin{equation}
D_n = \frac{1}{N_\mathrm{k}} \sum_{k=1}^{N_\mathrm{k}} 
\sqrt{\frac{\sum_{i=1}^{N_p}\left(\tilde{x}_{i,n}^{(k)} - \left\langle \tilde{x}_{i,n}^{(k)} \right\rangle\right)^2}{N_\mathrm{p}}},
\end{equation}
where $N_\mathrm{k}$ is the number of parameter kind and $N_\mathrm{p}$ is the population size. 
% $\left\langle\tilde{x}_{i,n}^{(k)}\right\rangle=\sum_i \tilde{x}_{i,n}^{(k)}/N_\mathrm{p}$ is the ensemble average. 
Given thresholds $\epsilon_{\mathrm{fit}}=0.005$ and $\epsilon_{\mathrm{div}}=0.10$, the tournament size $m$ is set by 
\begin{equation}
m =
\begin{cases}
3 ~(\Delta F_n < \epsilon_{\mathrm{fit}} ~\mathrm{and}~ D_n < \epsilon_{\mathrm{div}}) \\[6pt]
7 ~(\Delta F_n < \epsilon_{\mathrm{fit}} ~\mathrm{and}~ D_n > \epsilon_{\mathrm{div}}) \\[6pt]
5 ~(\Delta F_n > \epsilon_{\mathrm{fit}})
\end{cases}
\end{equation}
For example, when $\Delta F_n$ and $D_n$ are small, we regard the GA as having converged on a local optimum. 
In this case, we set the tournament size to 3 to increase the randomness and recover diversity.
\par
Crossover is an operator that creates two new individuals by exchanging part of the parameters of the parents in the Binary-coded GA.
However, in Real-coded GA, crossover is designed to generate children in the continuous search space built on the parameters of the parents. 
In this study, we adopt Simulated Binary Crossover (SBX) suggested in \citep{10.1162/106365601750190406}.
SBX is a Real-coded GA crossover method that incorporates elements of a single-point crossover from binary-coded GAs, where children are generated according to a probability distribution with peaks near the values of the parents. 
Two $n+1$-th generation children, $x^{(k)}_{c_1, n+1}$, and $x^{(k)}_{c_2, n+1}$, are generated from two parents in $n$-th generation $x^{(k)}_{p_1, n+1}$, and $x^{(k)}_{p_2, n+1}$ as
\begin{subequations} \label{eq:sbx_children}
\begin{align}
x^{(k)}_{c_1, n+1}
&=\frac{1}{2}\left[(1+\beta_\mathrm{CO})x^{(k)}_{p_1, n}+(1-\beta_\mathrm{CO})x^{(k)}_{p_2, n}\right], \\[1ex]
x^{(k)}_{c_2, n+1}
&=\frac{1}{2}\left[(1-\beta_\mathrm{CO})x^{(k)}_{p_1, n}+(1+\beta_\mathrm{CO})x^{(k)}_{p_2, n}\right].
\end{align}
\end{subequations}
$\beta_\mathrm{CO}$ determines the probability distribution 
and is given as
\begin{equation}
\beta_\mathrm{CO} =
\begin{cases}
(2u)^{\tfrac{1}{\eta_\mathrm{CO}+1}}, & \text{if } u \leq 0.5, \\[6pt]
\left[\dfrac{1}{2(1-u)}\right]^{\tfrac{1}{\eta_\mathrm{CO}+1}}, & \text{otherwise}.
\end{cases}
\end{equation}
%Shimizu 今更ですが、ηが被っている\eta_\mathrm{CO},ベータも
where $u$ is a uniform random number $(0\leq u \leq 1)$, $\eta_\mathrm{CO}$ is the distribution
index, which is a free nonnegative parameter. A large $\eta_\mathrm{CO}$ gives a high probability 
of being near the parents, and a small $\eta_\mathrm{CO}$ allows distant solutions from the parents. 
In this study, we set $\eta_\mathrm{CO}=2$. 
\par
Mutation is an operator that introduces random modifications to the part of parameters to maintain diversity and to avoid premature convergence to local optima. While crossover explores the search space by combining parental information, mutation plays a complementary role by perturbing individuals, thereby enhancing the exploration of unexplored regions. In this study, we adopt the Gaussian Mutation \citep[e.g.][]{beyer2002evolution}.

In the algorithm, the mutation occurs under the Gaussian probability function centered in parameter $x^{(k)}_{i,n}$, and whose width $\sigma$ is 10\% of the absolute value of the parameter, i.e., $\sigma=0.1\left|x^{(k)}_{i,n}\right|$.
\par
We note that the crossover and mutation are performed with probabilities of 80\% and 30\%, respectively. If they are not performed, the selected individuals are passed without any modification.
%%%%%%%%%%%%%%%%%%%%%%%%%%%%%%%%%%%%%%%%%%%%%%%%%%%%%%%%%%%%%%%%%%%%%%%%%%%%%%%%%%%%%%%%%%%%%%%%%%%%%%%%%%%
% \subsection{Overall procedure} \label{section:Initial condition}
% The optimal solution is found by repeating the following procedure:
% \begin{enumerate}[itemsep=0pt, topsep=0pt]
% \item We run dynamo simulations with parameters,
% \item We compare the simulation results with the observed/simulated ground truth data.
% \item We optimize the parameters with the GA procedure described in \ref{section:GA procedures}.
% \end{enumerate}
% We can obtain an optimal solution through the procedure (Figure \ref{figure_1}f).
% \par
%%%%%%%%%%%%%%%%%%%%%%%%%%%%%%%%%%%%%%%%%%%%%%%%%%%%%%%%%%%%%%%%%%%%%%%%%%%%%%%%%%%%%%%%%%%%%%%%%%%%%%%%%%%
%初期条件に関して
\par
The initial conditions of our numerical simulation are also prepared in our GA procedure. We first carry out a regular simulation with $u_0=695~\mathrm{cm~s^{-1}}$ and $s_0=50~\mathrm{cm~s^{-1}}$. 
%Shimizu ここに速度と周期、s0と磁場の関係を書くのはどうか
We can reproduce the average solar cycle, characterized by approximately 166 sunspots at maximum and a period of 11 years. The initial condition of this simulation is an arbitrarily large-scale field given as
\begin{align}
A_\phi &= 
\begin{cases}
0~&\mathrm{for}~r<0.7R_\odot, \\
\displaystyle{\frac{\sin\theta}{r^2}}~&\mathrm{for}~\geq 0.7R_\odot,
\end{cases}
\\
B_\phi &=0.
\end{align}
\begin{figure}[t]
  \centering
  \includegraphics[width=\linewidth]{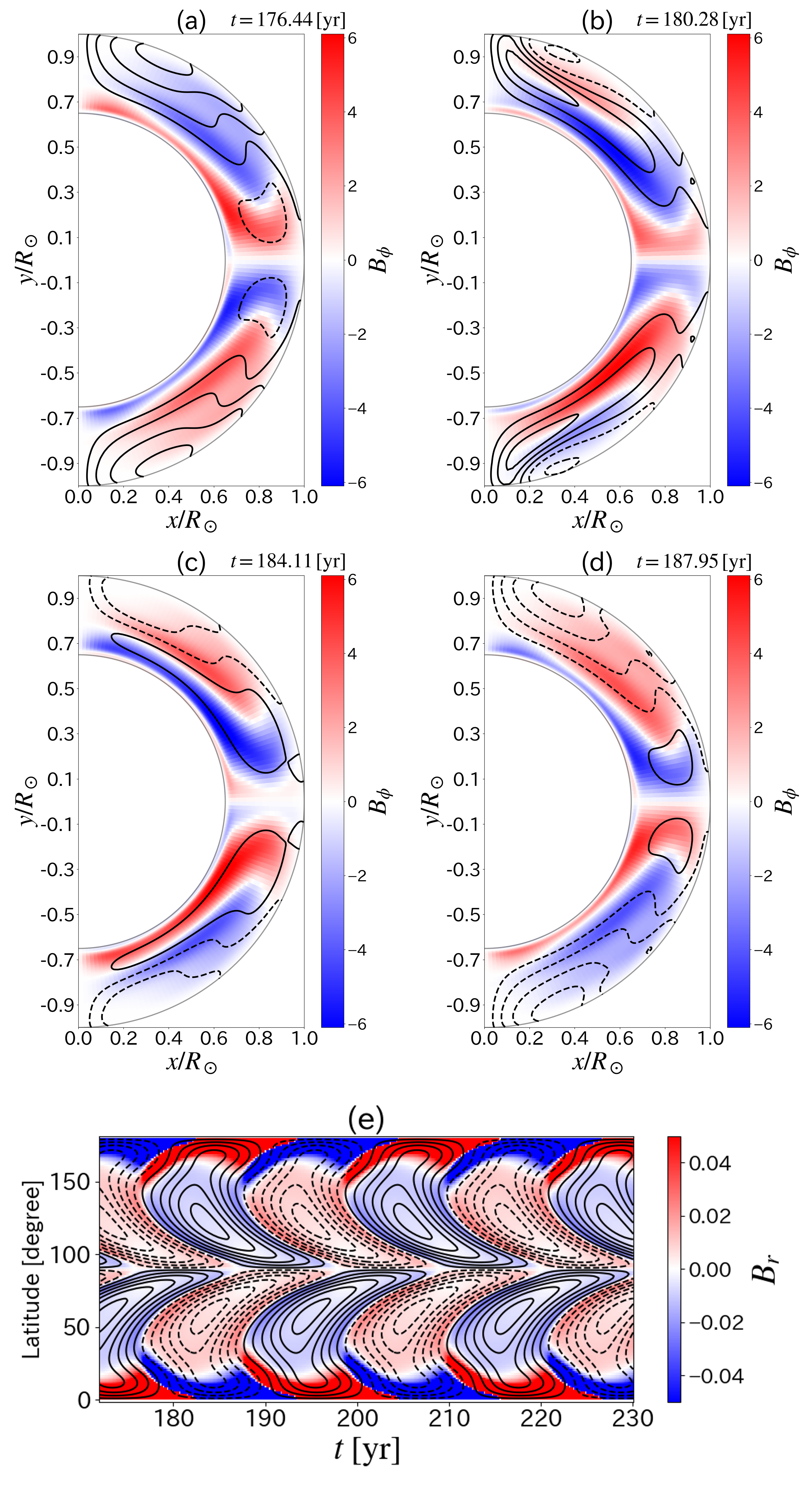}
  \caption{Evolution of the toroidal and poloidal magnetic field components. 
  Panels a, b, c, and d are results separated by intervals of 3.83 years.  
  In panels a-d, the red and blue colors represent the toroidal magnetic field $B_\phi$, and the contour lines are magnetic field lines of the poloidal field. The solid and dashed lines indicate clockwise and counterclockwise field.
  These results are characterized by a cycle period of $22.028~\mathrm{yr}$.
  Panel e shows a butterfly diagram for 58.2 years. The color contours represent $B_r$ at the top boundary ($r=R_\odot$), and the contour lines represent $B_\phi$ around the base of the convection zone $r=0.7R_\odot$.
  }
  \label{figure_2}
\end{figure}

% The value of the meridional flow ($u_0=695~\mathrm{cm~s^{-1}}$) is not necessarily realistic. 
% However, the significance of our results lies in the temporal evolution of $u_0$, 
% specifically whether it becomes larger or smaller compared with conventional values.
% Hotta: refereeに指摘されたら答えればいいことな気がしたので、ここでは書かないことにしました。
The simulation was run for 250 years, and representative results are shown in Figure \ref{figure_2}.
% Figure~\ref{figure_2}~(a),(b),(c) and (d) reveal how $B_\phi$ is amplified by the $\Omega$ effect at the base of the convection zone and how the magnetic field is transported by the meridional flow. Comparing (a) and (d) also shows that the polarity has reversed, resulting in nearly patterns.
Figure \ref{figure_2}e shows the butterfly diagram derived from simulation results. It exhibits characteristics and shapes consistent with observations, such as polarity reversing every 11 years and the polarity being opposite in the northern and southern hemispheres.
\par
Then, we use the data from the regular simulation with a cycle minimum as an initial condition and restart numerical simulations with parameters $u_0=a_\mathrm{s}^u$ and $s_0=a_\mathrm{s}^s$ obtained in the GA procedure. This simulation continues in a cycle minimum of around 80 years. The data in the cycle minimum is used as the initial condition for the main GA procedure.
Therefore, initial conditions vary between individuals, and the methodology employed in this study also explores optimal initial conditions. We note that due to the method of setting initial conditions, parameter inference always begins at a minimum.
%%%%%%%%%%%%%%%%%%%%%%%%%%%%%%%%%%%%%%%%%%%%%%%%%%%%%%%%%%%%%%%%%%%%%%%%%%%%%%%%%%%%%%%%%%%%%%%%%%%%%%%%%%%
\section{result}\label{sec:result}
% どんな結果を出したか

\subsection{Method validation with simulation generated data}\label{section:Validation of scheme}
\begin{figure}[t]
  \centering
  \includegraphics[width=\linewidth]{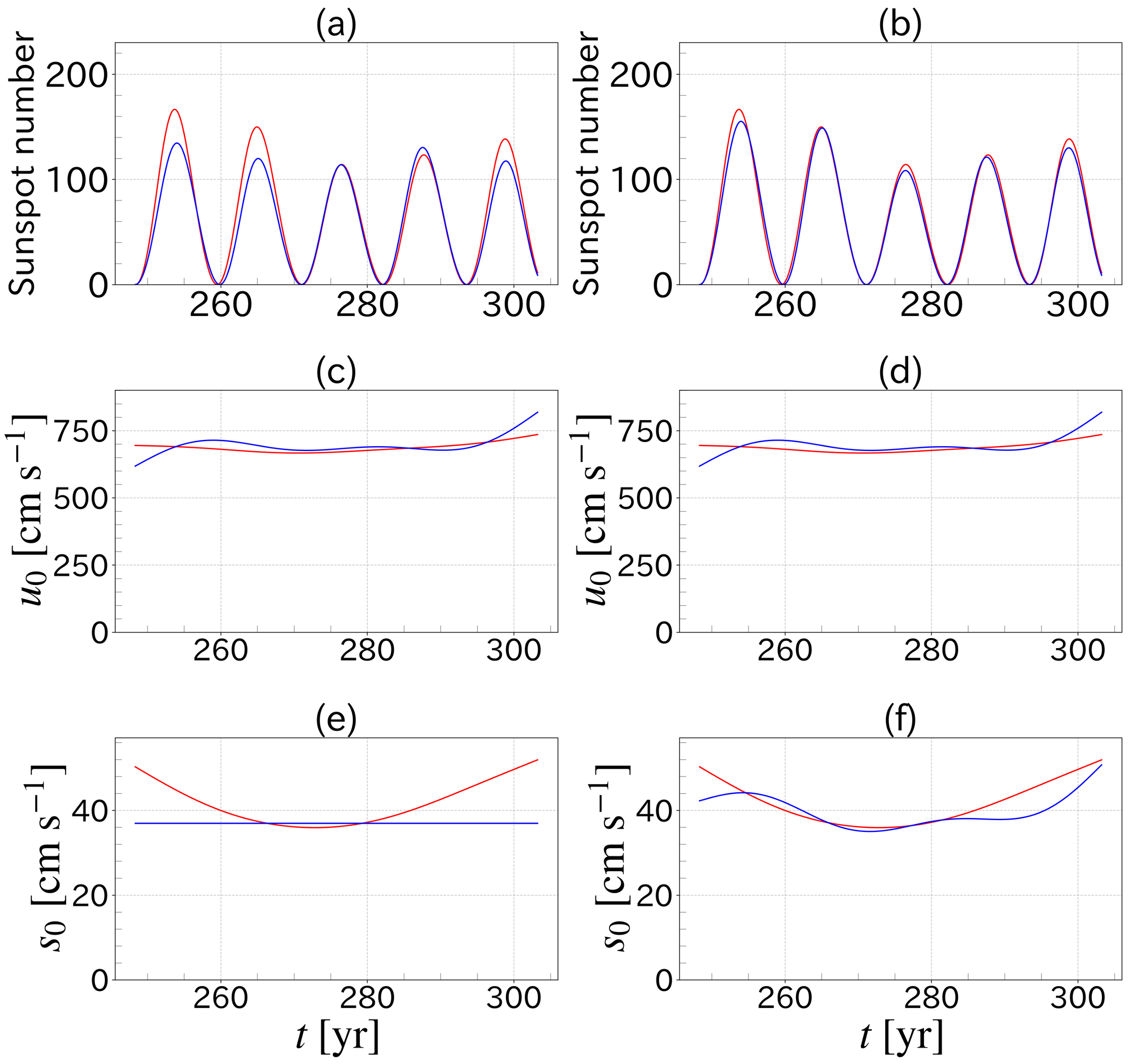}
  \caption{Comparison between the ground truth data generated by the simulation and our inference. Panels a, c, and e (Step 1) show the results of inferring only the time variation of $u_0$, and panels b, d, and f show the results of inferring the time variation of $s_0$ with $u_0$ obtained in Step 1. Panels a and b show the time variation of SSN.  Panels c and d show the time variation of $u_0$, and panels e and f show the time variation of $s_0$. The red and blue colors show the ground truth and our GA inference, respectively.
  }
  \label{figure_3}
\end{figure}
%TODO Dikpatiを引用したいが黒点数に関する言及が見つからない。
\begin{figure*}[t]
  \centering
  \includegraphics[width=\textwidth]{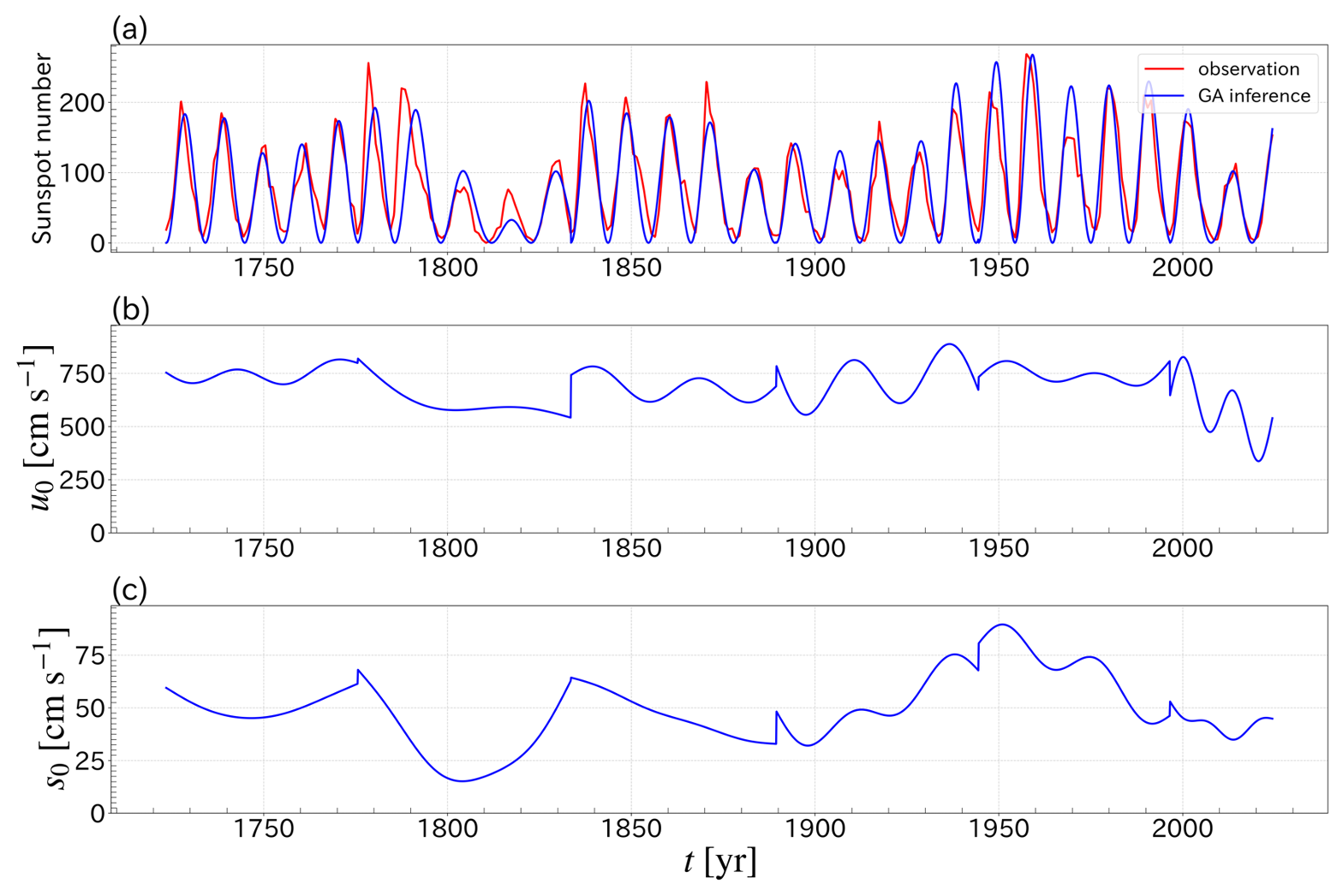}
  \caption{Inference result for the observational data. 
  Panel a shows the time variation of SSN. The red and blue represent the observational data and GA inference, respectively. The observed SSN is yearly mean total sunspot number by SILSO (Sunspot Number Version 2.0 \citep{silso_smoothed_ssn}.
  We interpolate the data to achieve 40-day intervals for the inference.
  Inference is performed every $5$ cycles, and we merge all the inferences into a plot. Panels b and c represent the time variations of the inferred $u_0$ and $s_0$, respectively.}
  \label{figure_4}
\end{figure*}

\begin{figure*}[t]
  \centering
  \includegraphics[width=.8\textwidth]{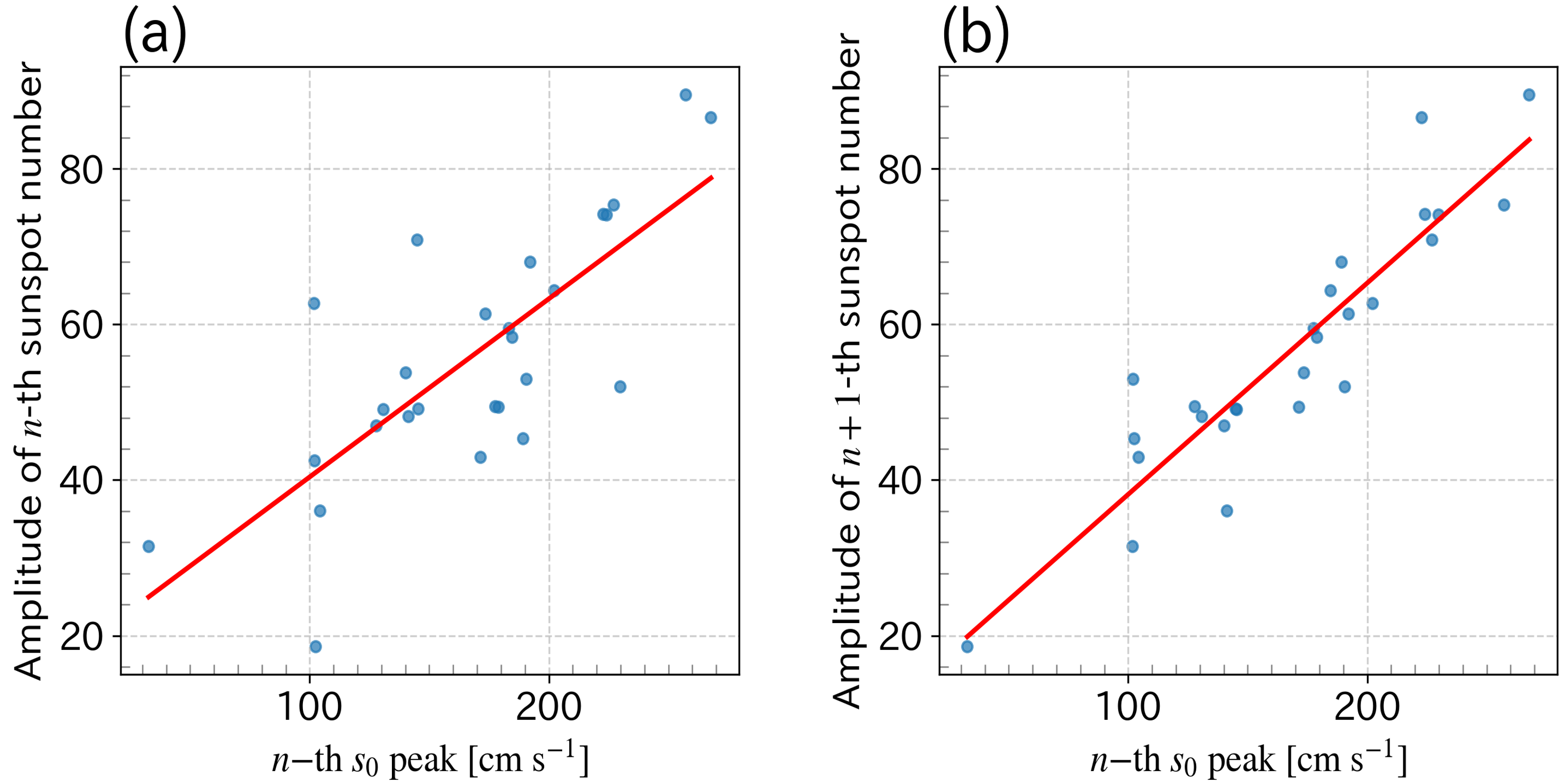}
  \caption{Relation between $s_0$ and sunspot numbers. Panel a shows a scatter plot of the peak value of $s_0$ at the $n$-th cycle and the amplitude of the $n$-th sunspot number. Panel b shows a scatter plot of the peak value of $s_0$ at the $n$-th cycle and the amplitude of the $n+1$-th sunspot number. The correlation coefficients for the panels a and b are 0.763 and 0.916, indicating a time lag between the poloidal and toroidal field generations.}
  \label{figure_add}
\end{figure*}

We evaluate the ability of our method by inferring the free parameters for ground truth data generated by numerical simulations in this section.
We can control the parameters $u_0$ and $s_0$ independently and evaluate the ability of our estimation.
The ground truth data is the simulation results for five cycles with both $u_0$ and $s_0$ varying over time. 
%mod1 We notice that optimization often falls into a local optimum when we infer parameters for meridional flow $u_0$ and $\alpha$ effect $s_0$ simultaneously. The meridional flow basically determines the cycle period. The faster meridional flow tends to cause a shorter solar cycle period, and vice versa.
The meridional flow basically determines the cycle period. The faster meridional flow tends to cause a shorter solar cycle period, and vice versa. 
Optimization often falls into a local optimum when we infer parameters for meridional flow $u_0$ and $\alpha$ effect $s_0$ simultaneously.
Thus, we infer them using the following procedure: 
\begin{description}
    \item[Step 1] Time dependent $u_0$ and fixed $s_0$ are inferred.
    \item[Step 2] Time dependent $s_0$ is inferred with $u_0$ obtained in Step 1.
\end{description}
Figures \ref{figure_3}a, c, and e show the results of Step 1. Figure \ref{figure_3}a shows the SSNs of the ground truth (red) and the GA inference (blue).  While the SSNs at maxima are slightly deviated, the overall structure can be reproduced. The amplitude of the meridional flow are recovered (Figure \ref{figure_3}c). While the amplitude of the $\alpha$-effect $s_0$ is constant over time in Step 1, the amplitude is roughly inferred. The correlation coefficient between the ground truth SSNs and the GA inference is $0.975$. The mean absolute percentage error for $u_0$ and $s_0$ are $2.87\%$ and $10.89\%$, respectively.
Figures \ref{figure_3}b, d, and f show the results of Step. 2. Compared with \ref{figure_3}a, Step 2 improves the reproduction ability, especially about the amplitude of the SSN (Figure \ref{figure_3}b). 
% The first cycle deviates more than that in Step 1. We expect this to occur because of inappropriate initial conditions for inference. 
The sunspot correlation coefficient in Step 2 is $0.995$, and the mean absolute percentage error for $s_0$ is 5.25\%, which is reduced from that in Step 1. According to Figure \ref{figure_3}f, the trend in the time variation of $s_0$ can be inferred. 
% Therefore, it is considered that $s_0$ can be sufficiently inferred the trend even in actual observational data.
\subsection{Application to observation data}

We apply our method to observational SSN data from 1723 to 2024 and perform inference of the dynamo parameters $u_0$ and $s_0$ (Figure \ref{figure_4}). As longer time inference is very challenging, we divide the SSN data into 5-cycle units. Inference is separately applied to each 5-cycle unit. Thus, the results are not continuous. 
It is preferable to infer as long duration as possible at once. We confirm that inferring with 3 or 4-cycle unit does work. While we expect that longer than 5-cycle unit is also possible, the difficulty of inference increases. As a compromise, we choose the 5-cycle unit.

We conducted the GA with a maximum of 50th generation. While we use the fitness parameter defined in equation (\ref{eq:fitness}) for most of the period, we employ a different definition for the period 
1775--1833, which corresponds to the Dalton Minimum. The mean square error
%mod1 correlation coefficient 
for the cycle period is used as the fitness parameter in Step 1. 
%mod1 , and only the SSN correlation coefficient $r_\mathrm{sunspot}$ is used as the fitness parameter in Step 2.
%Shimizu サイクルごとの周期の相関係数です。今回の場合5サイクル（1775–1887の中に）あるのでそれぞれのサイクルの長さから相関係数を出しています。
The different definition is used because the original fitness parameter in Step 1 attempts to reproduce the sharp peak fluctuations in the observed data solely through the time variation of $u_0$, leading to a local optimum. 
%mod1 step2は従来通りやったので以下不要
% \textcolor{red}{Furthermore, in Step 2, the inferred parameters for the minimum period around 1790 deviate from the observational data, so introducing $\mathrm{NMSE}$ ($e_\mathrm{sunspot}$ into the fitness parameter causes optimization to proceed in a direction diverging from the observations.}
\par
We first discuss the overall structure of the inference (Figure \ref{figure_4}a). The number of sunspots at the maximum and the periodicity are acceptably reproduced. Figure \ref{figure_4}b shows that $u_0$ fluctuates within the range of $-51.1\%$ to $+29.0\%$ of time average while $s_0$ exhibits a larger fluctuation range (Figure \ref{figure_4}c), with fluctuation of $-69.1\%$ to $+82.7\%$. 
We calculate the correlation coefficient between cycle rise time and the peak sunspot number to evaluate the ability of the GA inference in this study. The relation is known as the Waldmeier effect \citep{1935MiZur..14..105W, 1939MiZur..14..470W}. In our inferred data, the correlation coefficient is -0.526, indicating an acceptable relation between them. We note that the correlation coefficient evaluated with the observed sunspot number is -0.766.

\par
Then, we examine the results of each 5-cycle unit. The SSN correlation coefficients obtained in chronological order from the oldest data are 0.927, 0.685, 0.933, 0.935, 0.873, and 0.974. 
Figures \ref{figure_4}a and b clearly show that the meridional flow velocity decreased during the Dalton Minimum from $1790$ to $1830$.
\cite{2021NatSR..11.5482M} expected that the meridional flow decreased during a grand minimum \citep{2010ApJ...724.1021K}; our analysis can support this idea.
A comparison between Figures \ref{figure_4}a and c indicates that $s_0$ becomes smaller when the SSN is relatively low (around 1750, 1816, and 1905). It is worth noting that the timing of $s_0$ minimum (1746, 1804, and 1898) is slightly earlier than the timing of sunspot amplitude minimum (1750, 1816, and 1905).

\begin{figure*}[t]
  \centering
  \includegraphics[width=\textwidth]{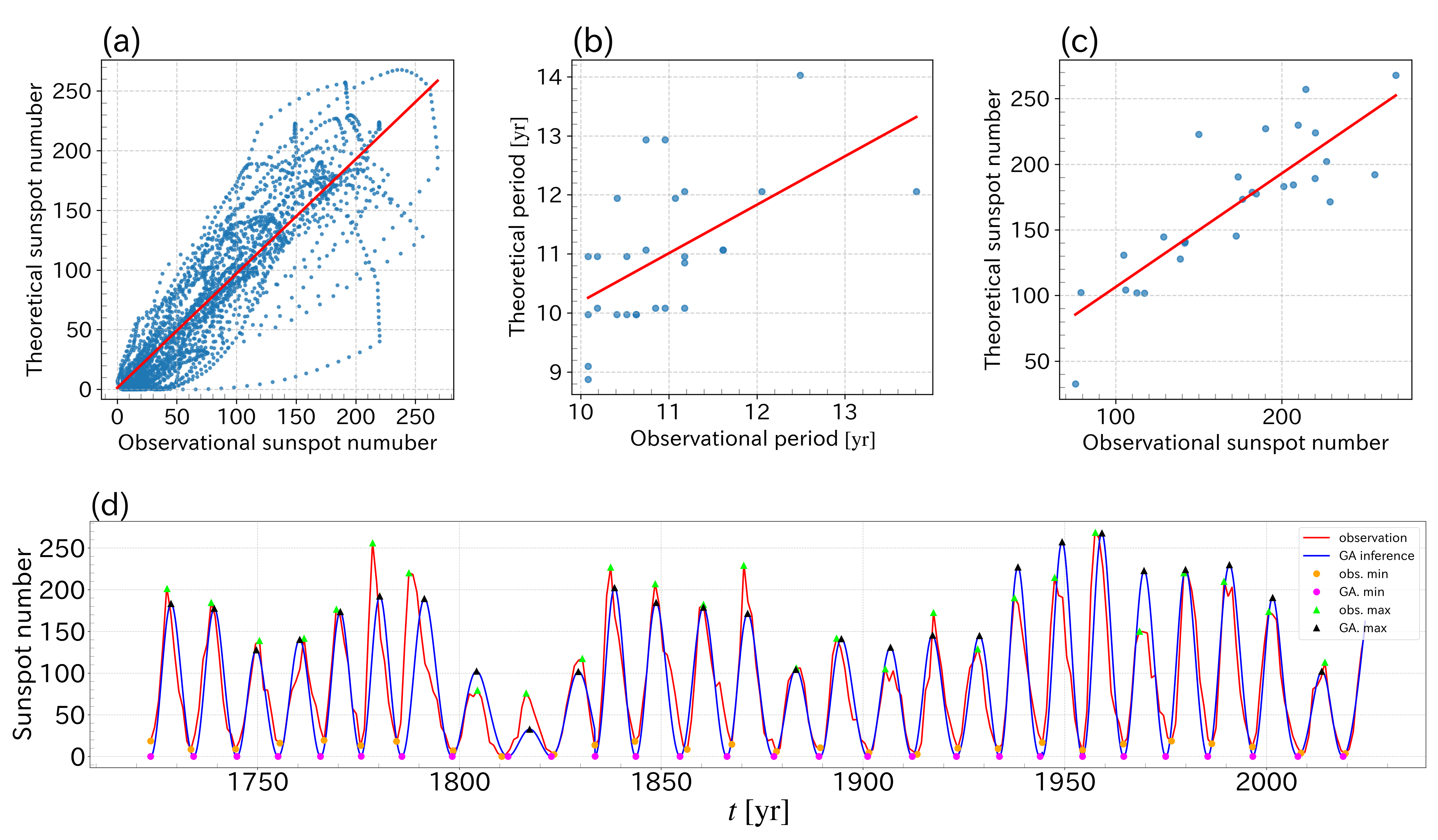}
  \caption{Statistical comparison of GA inference with observations. Panel a shows a scatter plot of SSN. The correlation coefficient is $0.876$. Panel b shows the scatter plot of the cycle periods. The correlation coefficient is $0.566$. The length of the cycle is measured by determining the minimum point, as indicated by the pink and orange circles in panel d. Panel c shows the correlation of SSN at the maximum period.  The correlation coefficient is $0.842$. The SSN at the maximum period is indicated by light green and black triangles for the observation and the GA inference, respectively.}
  \label{figure_5}
\end{figure*}

There should be a time lag between the $\alpha$ effect, creating a poloidal magnetic field on the solar surface, and toroidal field generation around the base of the convection zone by the $\Omega$ effect. 
We evaluate two correlation coefficients. One is that between the peak value of $s_0$ at the $n$-th cycle and the amplitude of the $n$-th sunspot number (Figure \ref{figure_add}a). The other is that between the peak value of $s_0$ at the $n$-th cycle and the amplitude of the $n+1$-th sunspot number (Figure \ref{figure_add}b). The correlation coefficient for the former and the latter is 0.763 and 0.916, respectively. The better coefficient of the latter partly implies the time lag between the poloidal and toroidal field generations.
The somewhat poor correlation value of 0.763 in Figure \ref{figure_add}a implies that the solar cycle memory is disturbed by the fluctuations and the toroidal cycle amplitude to poloidal as determined by $s_0$ value in Figure \ref{figure_add}a. It is not strongly correlated as expected for the Sun.
% {Regarding the time lag, we investigated the correlation between the peak value of $s_0$ at the $n$-th cycle and the amplitude of the $n$-th sunspot number(figure \ref{figure_add}(a)) or the amplitude of the $n+1$-th sunspot number (figure \ref{figure_add}(b)) .The correlation coefficient of (a) is 0.763, whereas that of (b) is 0.916. Therefore, we can say that a tendency for the time lag becomes apparent.}

\par
We observe that the amplitude for cycle 20 (1964--1976) has not been precisely reproduced. 
\cite{Dey_2025} suggest that the cycle 20's weakness is due to the reduced tilt angle of the sunspots for cycle 19, which generated a weak polar field and the weak following cycle. The unsuccessful reproduction in our study is thought to stem from an inability to estimate the abrupt change in $s_0$ during cycle 19.
\par
Figure \ref{figure_5} presents several statistical evaluations of the overall inference results.
%Simizu fig5(c)の説明がなかったので追加しました
Figure \ref{figure_5}a shows a scatter plot for all corresponding SSN,
with a correlation coefficient of 0.876. 
Data points significantly deviating from the regression line to the lower right correspond to the results around 1784--1798. Figure \ref{figure_5}b shows the scatter plot for each cycle period, with a correlation coefficient of $0.566$. This value is worse than that in \cite{2010ApJ...724.1021K} because our optimization does not include the cycle period as a fitness parameter. We intend to optimize the free parameters to reproduce the overall structure of the SSN rather than solely the cycle period. 
Figure \ref{figure_5}c shows a scatter plot of SSN at each maximum, 
with a correlation coefficient of 0.842. This value is superior to that in \cite{2010ApJ...724.1021K} because we also optimize $s_0$.

%%%%%%%%%%%%%%%%%%%%%%%%%%%%%%%%%%%%%%%%%%%%%%%%%%%%%%%%%%%%%%%%%%%%%%%%%%%%%%%%%%%%%%
\section{Summary and discussion}
We develop a method to reproduce the observed solar cycles in numerical simulations, utilizing automatic inference of free parameters. We try to infer the amplitudes of the meridional flow $u_0$ and the $\alpha$ effect. The temporal evolutions of these are also inferred. These parameters are expanded (Equation \eqref{eq:time_variation}) with the sine functions, and the coefficients for these sine functions are inferred. As an inference method, we adopt a GA method. We carry out a number of simulations and evaluate their fitness against ground truth data with certain parameters. Then, the following generation is created using ``Selection,'' ``Crossover,'' and ``Mutation.'' As the generation progresses, the ability of numerical simulations to replicate the observation improves.
We apply this method to observational data (1723-2024) and successfully reproduce the overall structure in the time series of SSN.
Regarding the time lag between the poloidal and toroidal field generations confirmed from the reproduction and inference results, we can consider exploring the memory of the solar cycle predicted in \cite{2021ApJ...913...65K} for the future work.
Several issues remain to be addressed in future studies. 
The fitness parameter is open to discussion. 
For example, we consider that directly incorporating the period or the position of the cycle minima into the fitness will improve the inference accuracy of $u_0$ and enhance the accuracy of period reproduction.
We intend to reproduce the overall structure, but there is a possibility of using some representative value, such as the SSN, in the maximum and the cycle period. Furthermore, we just use the SSN in this study, but our algorithm can be extended to reproduce the global magnetic field observed in modern instrumentation. The choice of the initial condition is also possibly improved. We generate the initial condition for the 5-cycle unit independently. We can inherit the final state of a 5-cycle unit magnetic field to the next unit. This allows us to obtain smooth, continuous results for SSN, $u_0$, and $s_0$, which we believe brings the model closer to reality. \par
As a further application, we plan to apply our method to the SSN obtained by isotope analysis, such as $^{14}\mathrm{C}$ \citep{2004SoPh..224..317M} and 
$^{10}\mathrm{Be}$ \citep{2004A&A...413..745U}
%Shimizu please refer paper(s) about 10Be, maybe Usoskin et al.,
 and historical documents \citep{2025arXiv250214665H}.
% Given the customizable nature of this approach, we anticipate that our scheme will be effectively utilized for a variety of purposes.

% TODO セクションで始める

% \begin{figure}[t]
%   \centering
%   \includegraphics[width=\linewidth]{AAJ/figure_2.png}
%   \caption{Evolution of the toroidal and poloidal magnetic field components. 
%   Panels a, b, c, and d are results separated by intervals of 3.83 years.  
%   In panels a-d, the red and blue colors represent the toroidal magnetic field $B_\phi$, and the contour lines are magnetic field lines of the poloidal field. The solid and dashed lines indicate clockwise and counterclockwise field.
%   These results are characterized by a cycle period of $22.028~\mathrm{yr}$.
%   Panel e shows a butterfly diagram for 58.2 years. The color contours represent $B_r$ at the top boundary ($r=R_\odot$), and the contour lines represent $B_\phi$ around the base of the convection zone $r=0.7R_\odot$.
%   }
%   \label{figure_2}
% \end{figure}
\par
\begin{acknowledgments}
H.H. is supported by JSPS KAKENHI grant Nos. JP25H00673, JP25K22031, JP23K25906, JP21H04497, and JP21H04492 and MEXT as a Programme for Promoting Researchers on the Supercomputer Fugaku (JPMXP1020230504).
\end{acknowledgments}

\appendix
\section{Numerical method}
We describe the second-order Total Variation Diminishing (TVD) Runge-Kutta method \citep{SHU1988439}. 
We follow the procedure shown below to advance a time step. We use an arbitrary variable $\phi$ and we assume to solve an equation as 
\begin{align}
    \frac{\partial \phi}{\partial t} = R(\phi),
\end{align}
where $R$ is an arbitrary operator.
First, we use the forward Euler method to derive $\phi^{(a)}$ and $\phi^{(b)}$ from $n$-th step value $\phi^n$. $\phi^{(a)}$ and $\phi^{(b)}$ are given as
\begin{subequations}
\begin{align}
    \phi^{(a)} &= \phi^n + \Delta tR(\phi^n), \\[1ex]
    \phi^{(b)} &= \phi^{(a)} + \Delta t R(\phi^{(a)}).
\end{align}
\end{subequations}
Second, we calculate $\phi^{n+1}$ by taking the average of the $\phi^{(b)}$ and $\phi^n$ values obtained. $\phi^{n+1}$ is given as
\begin{equation}
    \phi^{n+1} = \frac12\phi^n + \frac12\phi^{(b)}.
\end{equation}
Using this scheme, equation\eqref{eq:induction} can be solved with satisfying the TVD condition.

\bibliography{sample701}
\bibliographystyle{aasjournalv7}


\end{document}